# An evaluation of empirical equations for assessing local scour around bridge piers using global sensitivity analysis

**Gianna Gavriel**
Postdoctoral Researcher
School of Engineering
University of Galway, University Road, Galway, Ireland, H91 TK33
and
School of Civil, Aerospace and Design Engineering
Faculty of Science and Engineering
University of Bristol, Queen's Building, University Walk, Bristol, BS8 1TR, UK
email: gianna.gavriel@universityofgalway.ie
ORCID: https://orcid.org/0000-0002-4648-1706

**Maria Pregnolato**
Associate Professor in Flood Risk & Resilient Infrastructure
Department of Hydraulic Engineering
Faculty of Civil Engineering and Geoscience
Delft University of Technology, Stevinweg 1, Delft, 2628CN, the Netherlands
and
School of Civil, Aerospace and Design Engineering
Faculty of Science and Engineering
University of Bristol, Queen's Building, University Walk, Bristol, BS8 1TR, UK
email: m.pregnolato@tudelft.nl
ORCID: https://orcid.org/0000-0003-0796-9618

**Francesca Pianosi**
Associate Professor in Water & Environmental Engineering
School of Civil, Aerospace and Design Engineering
Faculty of Science and Engineering
University of Bristol, Queen's Building, University Walk, Bristol, BS8 1TR, UK
email: francesca.pianosi@bristol.ac.uk
ORCID: https://orcid.org/0000-0002-1516-2163

**Theo Tryfonas**
Professor of Infrastructure Systems and Urban Innovation
School of Civil, Aerospace and Design Engineering
Faculty of Science and Engineering
University of Bristol, Queen's Building, University Walk, Bristol, BS8 1TR, UK
email: theo.tryfonas@bristol.ac.uk
ORCID: https://orcid.org/0000-0003-4024-8003

**Paul J. Vardanega**
Associate Professor in Civil Engineering
School of Civil, Aerospace and Design Engineering
Faculty of Science and Engineering
University of Bristol, Queen's Building, University Walk, Bristol, BS8 1TR, UK
email: p.j.vardanega@bristol.ac.uk (Corresponding Author)
ORCID: https://orcid.org/0000-0001-7177-7851

Date of version: 12$^{th}$ January 2026

# An evaluation of empirical equations for assessing local scour around bridge piers using global sensitivity analysis

G. Gavriel, M. Pregnolato, F. Pianosi, T. Tryfonas and P.J.Vardanega

**Abstract**

Bridge scour is a complex phenomenon combining hydrological, geotechnical and structural processes. Bridge scour is the leading cause of bridge collapse, which can bring catastrophic consequences including the loss of life. Estimating scour on bridges is an important task for engineers assessing bridge system performance. Overestimation of scour depths during design may lead to excess spendings on construction whereas underestimation can lead to the collapse of a bridge. Many empirical equations have been developed over the years to assess scour depth at bridge piers. These equations have only been calibrated with laboratory data or very few field data. This paper compares eight equations including the *UK CIRIA C742 approach* to establish their accuracy using the open access USGS pier-scour database for both field and laboratory conditions. A one-at-the-time sensitivity assessment and a global sensitivity analysis were then applied to identify the most significant parameters in the eight scour equations. The paper shows that using a global approach, i.e. one where all parameters are varied simultaneously, provides more insights than a traditional one-at-the-time approach. The main findings are that the *CIRIA* and *Froehlich* equations are the most accurate equations for field conditions, and that angle of attack, pier shape and the approach flow depth are the most influential parameters. Efforts to reduce uncertainty of these three parameters would maximise increase of scour estimate precision.

**List of Symbols and Abbreviations**

<u>Latin symbols</u>

| Notation | Parameter | Units |
|---|---|---|
| $B$ | Pier width | $m$ |
| $B'$ | Effective pier width | |
| $D_{50}$ | Medium soil bed gradation | $mm$ |
| $F_{(pier)}$ | Froude number for the *TAMU* equation | - |
| $F_a$ | Froude number for the Froehlich equation | - |
| $F_{c(pier)}$ | Froude number for critical velocity for the TAMU equation | - |
| $F_r$ | Froude number for the *Chitale* equation | - |
| $F_{r1}$ | Froude number for the *HEC-18* equation | - |
| $F_y$ | Unconditional Cumulative Distribution Function for the PAWN analysis | - |
| $g$ | Gravitational acceleration | $m/s^2$ |
| $k$ | Number of parameters in the equation of the distribution fitted to a database for estimating the Akaike statistic | - |
| $K_1$ | Pier nose geometry correction factor for the *HEC-18* equation | - |
| $K_2$ | Bed condition correction factor for the *HEC-18* equation | - |
| $K_3$ | Attack angle correction factor for the *HEC-18* equation | - |
| $K_a$ | Pier alignment influence factor for the *Melville* and *Melville & Sutherland* equations | - |
| $K_{aL}$ | Correction factor for the effect of the attack angle for the *Laursen* equation. | - |
| $K_{angle}$ | Correction factor for the effect of the attack angle for the *CIRIA* equation | - |
| $K_d$ | Correction factor for the effect of medium sediment size for the *Melville & Sutherland* equation | - |
| $K_{depth}$ | Correction factor for the effect of the water depth for the *CIRIA* equation | - |
| $K_{h4}$ | Sediment gradation and size influence factor for the *Melville* equation | - |
| $K_I$ | Intensity influence factor for the *Melville* and *Melville & Sutherland* equations | - |
| $K_{pa}$ | Aspect ratio influence factor for the *TAMU* equation | - |
| $K_{psh}$ | Pier shape influence factor for the *TAMU* equation | - |
| $K_{psp}$ | Pier spacing influence factor for the *TAMU* equation | - |
| $K_{pw}$ | Water depth influence factor for the *TAMU* equation | - |
| $K_s$ | Pier sediment size influence factor for the *Melville* and *Melville & Sutherland* equations | - |
| $K_{sh}$ | Correction factor for the effect of pier nose shape for the *Laursen* equation | - |
| $K_{shape}$ | Correction factor for the effect of the attack angle for the *CIRIA* equation | - |
| $K_{velocity}$ | Correction factor for the effect of the water flow velocity for the *CIRIA* equation | - |
| $K_y$ | Flow intensity influence factor for the *Melville* and *Melville & Sutherland* equations | - |
| $K_\theta$ | Angle of attack influence factor for *Melville* | - |
| $K_\sigma$ | Pier shape influence factor for the *Melville* and *Melville & Sutherland* equations | - |
| $L$ | Pier length | m |
| $n$ | Number of input parameters | - |
| $N$ | Number of generated outputs from the *PAWN* method | - |
| $S$ | Spacing between the centers of two piers. | m |
| $Sh$ | Bridge pier nose shape | - |
| $t_{s(\Delta)}$ | Maximum difference between the maximum and minimum value of the coefficient of variation for one-at-a-time sensitivity analysis | - |
| $V_1$ | Approach velocity | m/s |
| $V_c$ | Critical velocity | m/s |
| $y_1$ | Flow depth | m |
| $y_{bc}$ | Calculated scour depth based on the base case scenario for one-at-a-time sensitivity analysis | m |
| $y_c$ | Calculated scour depth for one-at-a-time sensitivity analysis based on the average values of input parameters | m/s |
| $y_s$ | Scour depth | m |

<u>Greek Symbols</u>

| **Notation** | **Parameter** | **Units** |
|---|---|---|
| $\theta$ | Attack angle | ° |
| $\lambda$ | Rate parameter for estimating log-likelihood | - |
| $\mu$ | Mean | - |
| $\pi$ | Value of pi | - |
| $\sigma$ | Standard deviation | - |
| $\varphi$ | Attack angle influence factor for the *Froehlich* equation | ° |

Statistical Terms

| Notation | Definition |
|---|---|
| $t_s$ | Spread for the one-at-a-time sensitivity analysis |
| $KS(I_k) = \max |F_y(y) - F_{y|x_i}(y|x_i \in I_k)|$ | Kolmogotov-Smirnov statistic |
| $f_{(x_i)}$ | Calculated scour depth for the one-at-a-time sensitivity analysis |
| $f_{(x_b)}$ | Calculated scour depth for the base case scenario |
| $S_i = \underset{k=1,\ldots,n}{\mathrm{mean}}\, KS(I_k)$ | PAWN sensitivity index |

Abbreviations

| Acronym | Definition |
|---|---|
| GSA | Global Sensitivity Analysis |
| OAT | One-at-a-time sensitivity analysis |
| USGS | US Geological Survey |
| Gam | Gamma Distribution |
| GEV | General Extreme Value Distribution |
| LogNorm | Normal Logarithmic Distribution |
| Uni | Uniform Distribution |

## 1. Introduction

Bridge scour is the removal of material (erosion) caused by flowing water and wave action around bridge piers, abutments and foundations (e.g., Melville & Sutherland 1988; Melville 1997; Graf & Istiarto 2002; Arenson et al. 2012; Briaud et al. 2014; Kirby et al. 2015; Qi et al. 2016). If the depth of scour reaches a significant extent, it can jeopardize the stability of the bridge foundations, posing a risk to the structure which may result in damage or collapse (e.g., Kirby et al. 2015; Liang et al. 2015; Shahriar et al. 2021b). Scour around bridges is a common cause of bridge failure (e.g., Melville & Coleman 2000; Deng & Cai 2010; Maddison 2012; Arneson et al. 2012; Liang et al. 2015; Kirby et al. 2015; Dikanski et al. 2018; Ekuje 2020). Bridge scour is challenging for engineers to model and predict as it is a complex process involving multiple factors: the properties of the soil (geotechnical), the design of the bridge (structural), and the effects of water flow (hydraulic) (cf. National Academies of Sciences, Engineering, and Medicine 2011a; Dehghani et al. 2013).

According to Lauchlan & May (2002), the *Lacey equation* from the 1930s is one of the earliest recorded scour equations for calculating scour depth. Laursen & Toch (1956) used flume test data to investigate scour, however they were only able to investigate the influence of the angle of attack on bridge scour due to the technological limitations of the time (cf. Ettema et al. 2017). Many other empirical equations for estimation of scour depth have been developed (e.g., Melville & Sutherland 1988; Froehlich 1988; Melville 1997; Gaudio et al. 2010; Hong et al. 2012; Brandimarte et al. 2012; Arneson et al. 2012; Kirby et al. 2015; Briaud 2015a, 2015b; Qi et al. 2016; Yazdanfar et al. 2021; Kazemian et al. 2023).

Some formulae estimate the scour depth by considering the time taken for the scour hole to develop (Nandi & Das, 2023), however measuring the time for scour to develop is difficult to ascertain in field conditions (e.g., Melville & Chiew 1999, National Academies of Sciences, Engineering, and Medicine 2011b, Choi & Choi 2016). Recent studies using energy conservation have been reported by Sun et al. (2023) and approaches making use of Artificial Intelligence (AI) have been reported (e.g., Kumar et al. 2024).

Understanding the accuracy of available empirical (and semi-empirical) equations for estimating scour depth as well as the influential parameter(s) in said equations is important when using estimates from such equations in design and planning activities (cf. Arneson at al. 2012, Kirby et al. 2015). Previous studies comparing scour depth estimation equations have been limited in some cases by relatively small datasets. Most comparisons have not fully utilized the full USGS open-source database (Benedict & Caldwell 2014a, 2014b – see Section 2.1 for more details on this database) (cf. Table S2). Additionally, sensitivity analyses have largely been limited to one-at-a-time (OAT) approaches (cf. Table S2). These limitations have arguably left knowledge gaps in understanding how well empirical scour equations can predict scour depth across a range of field conditions and importantly which parameters in these equations have the most influence on the predicted scour depth values. This study addresses these gaps by leveraging the full USGS dataset and by applying Global Sensitivity Analysis (GSA) to provide a more comprehensive evaluation of a set of published equations for scour depth prediction.

Preliminary studies on the *USGS Database* have already been published by some of the authors of the present paper and these studies are listed in (Table 2). Gavriel et al. (2022) examined a small subset of the *USGS Database* (field data) to assesses the accuracy of two versions of the *HEC-18* equation (Richardson & Davis 2001; Arneson et al. 2012) and the *TAMU* equation (Briaud 2015a). To assess potential influences on the accuracy of the *HEC-18* (Arneson et al. 2012), Gavriel et al. (2023a) divided the *USGS Database* (field data) into three sets of subgroups: (i) scour depths of any value and scour depths below 2m; (ii) soil bed average particle size ranges of $D_{50} \leq 0.99$mm, $0.99$mm $< D_{50} \leq 9.99$mm, $D_{50} > 9.99$mm and (iii) by the field scour measurement technique employed. Gavriel et al. (2023b) applied a OAT sensitivity analysis to the *HEC-18* equation (again using the *USGS database*) to evaluate the sensitivity of the input parameters on the estimated scour depths.

This paper compares the accuracy of eight empirical time-independent equations when used to predict the measured scour depths for the laboratory and field datasets from the *USGS* *database* (some of the studied equations may be described as 'semi-empirical' but will be referred to as empirical for the remainder of this paper) (Benedict & Caldwell 2014a, 2014b). The eight equations are listed in Table 1 and include the *CIRIA* equation from the UK (Kirby et al. 2015); equations requiring the same parameters as CIRIA (i.e. the *Melville*, *Froehlich* and *Melville & Sutherland* equations) are also included. The *HEC-18* equation is included as it is an important equation from the USA (cf. Arenson et al. 2012), whose accuracy and sensitivity were previously analysed in Gavriel et al (2022, 2023a, 2023b), and one of the most often examined equations in the studies listed in Table S1. The *TAMU* equation from Briaud (2015a) was included as, unlike the other equations, it also uses the spacing between piers to calculate the maximum scour depth (Briaud 2015a). *Laursen* and *Chitale* are two older equations and were included as examples of equations with less parameters. All equations listed in Table 1 except *Froehlich* were originally calibrated either partly or entirely using laboratory data (cf. Briaud 2015a, Kirby et al. 2015, Arneson et al. 2012, Melville 1997, Froehlich 1988, Melville & Sutherland 1988, Blench et al. 1962, Laursen 1962).

The paper then applies an OAT sensitivity analysis, whereby each parameter is varied from its default value ('baseline') while keeping the others fixed to their default. While such OAT approaches are widely applied, it has two major limitations (Saltelli et al. 2019): (i) OAT sensitivity analysis does not capture interactions between parameters that may amplify or dampen the effects of individual variations – a common feature of non-linear models (cf. Saltelli et al. 2019) and (ii) the results are conditional on the chosen baseline, leaving the question open of how different the sensitivity assessment would be if a different 'baseline' was selected (cf. Saltelli et al. 2019). To overcome these limitations, a GSA (Pianosi et al. 2016) approach was also used in this work (to the authors' knowledge for the first time in the field of bridge scour) whereby all parameters are let vary simultaneously and the sensitivity is evaluated by the changes in the output distributions induced by the aforesaid simultaneous variations. For more details on the study presented in this paper the reader is directed to the doctoral thesis of Gavriel (2025).

**Table 1: The eight equations for bridge pier scour depth to be assessed in this study**

| Source | Name | Equation | Data used for calibration |
|---|---|---|---|
| Kirby et al. (2015) | *CIRIA* | $y_s = B K_{shape} K_{depth} K_{velocity} K_{angle}$ | Field, laboratory |
| Briaud (2015a) | *TAMU* | $y_s = 2.2 B' K_{pw} K_{psh} K_{pa} K_{psp} (2.6 F_{(pier)} - F_{c(pier)})^{0.7}$ | Laboratory |
| Arneson et al. (2012) | *HEC-18* | $y_s = 2.0 y_1 K_1 K_2 K_3 \left(\dfrac{B}{y_1}\right)^{0.65} Fr_1^{0.43}$ | Laboratory |
| Melville (1997) | *Melville* | $y_s = K_s K_\theta K_I k_{yb} k_G k_d$ | Laboratory |
| Froehlich (1988) | *Froehlich* | $y_s = 0.32 B\varphi \left(\dfrac{B'}{B}\right)^{0.62} \left(\dfrac{y_1}{B}\right)^{0.46} Fr_1^{0.20} \left(\dfrac{B}{D_{50}}\right)^{0.08}$ | Field |
| Melville & Sutherland (1988) | *Melville & Sutherland* | $y_s = B K_s K_a K_I K_y K_d K_\sigma$ | Laboratory |
| Chitale writing in Blench et al. (1962) | *Chitale* | $y_s = y_1(-0.51 + 6.65 F_r - 5.49 F_r^2) + y_1$ | Laboratory |
| Laursen (1962) | *Laursen* | $y_s = 1.5 B \left(\dfrac{y_1}{B}\right)^{0.3} K_{aL} K_{sh}$ | Laboratory |

**Table 2: Preliminary research studies related to this work**

| Reference | Methods Studied | Data | Methodology | Summary Findings |
|---|---|---|---|---|
| Gavriel et al. (2022) | *HEC-18*, *TAMU* | 19 USGS field data (from Maine, USA) | - Plots of measured versus calculated scour depth | - General trend of overestimation of pier scour depth observed (for the limited data set examined) |
| Gavriel et al. (2023a) | *HEC-18* | 936 USGS field data | - Plots of measured versus calculated scour depth<br>- Subsets: (i) scour depth of any value and scour depths below 2m (ii) $D_{50} \leq 0.99$mm, $0.99$mm $< D_{50} \leq 9.99$mm, $D_{50} > 9.99$mm and (iii) type of apparatus used for scour depth measurement | - The *HEC-18* equation provided a better prediction for lower values of scour depth |
| Gavriel et al. (2023b) | *HEC-18* | 936 USGS field data | - One-at-a-time sensitivity analysis<br>- Base Case 1 (BC1): all input parameters equal to their mean value<br>- Base Case 2 (BC2): all input parameters equal to their mean value except the angle of attack where $\theta = 0$ | - BC1 (attack angle of zero): pier width most influential parameter followed by flow velocity.<br>- BC2 (attach angle non-zero): attack angle and the pier width highly influential |
| Gavriel et al. (2025) | *HEC-18, TAMU, CIRIA, Melville, Melville & Sutherland, Froehlich, Chitale, Laursen* | 936 USGS field data | - Plots of measured versus calculated scour depth<br>- Subset: type of apparatus used for scour depth measurement<br>- Applied the Vardanega et al. (2021) framework to the scour measurement approaches from the *USGS Database* | - Fathometer Soundings, the most popular device in the *USGS Database*, was assessed as 'Moderately Applicable' according to the Vardanega et al. (2021) framework |

**Table 3: Parameter ranges used for the one-at-a-time sensitivity analysis (OAT) and Global Sensitivity Analysis (GSA)**

| Parameter | Units | Definition | Field Data (n=936) | | | | | Laboratory Data (n=569) | | | | |
|---|---|---|---|---|---|---|---|---|---|---|---|---|
| | | | OAT | | | | GSA | OAT | | | | GSA |
| | | | min | max | $\mu$ | $\sigma$ | Distribution | min | max | $\mu$ | $\sigma$ | Distribution |
| $B$ | m | Pier Width | 0.29 | 22.86 | 1.23 | 1.14 | GEV [0.26, 0.82, 0.42] | 0.02 | 0.92 | 0.11 | 0.14 | GEV [0.50, 0.06, 0.032] |
| $L$ | m | Pier Length | 2.44 | 74.88 | 11.15 | 5.49 | GEV [0.015, 9.18, 3.12] | - | - | - | - | - |
| $y_1$ | m | Approach Flow Depth | 0.10 | 22.52 | 3.96 | 3.07 | GEV [0.23, 2.38, 1.87] | 0.02 | 1.90 | 0.27 | 0.24 | GEV [0.47, 0.15, 0.10] |
| $V_1$ | m/s | Approach Flow Velocity | 0.02 | 4.48 | 1.26 | 0.78 | Gam [2.23, 0.71] | 0.15 | 2.16 | 0.51 | 0.32 | GEV [0.40, 0.34, 0.16] |
| $V_c$ | m/s | Critical Velocity | - | - | - | - | - | 0.22 | 1.27 | 0.44 | 0.22 | GEV [0.50, 0.33, 0.089] |
| $\theta$ | ° | Attack Angle | 0 | 85.00 | 19.91 | 17.55 | GEV [0.37, 11.18, 8.36] | - | - | - | - | - |
| $D_{50}$ | mm | Soil-bed Gradation | 0.01 | 108.00 | 19.08 | 28.04 | LogNorm [1.25, 2.12] | 0.22 | 7.80 | 1.23 | 1.41 | GEV [0.85, 0.50, 0.38] |
| $Sh$ | - | Pier Nose Shape | - | - | - | - | Uni [0.9, 2] | - | - | - | - | Uni [0.9, 2] |
| $S$ | m | Pier Spacing | - | - | - | - | Uni [1, 30] | - | - | - | - | Uni [1, 5] |

[**]$y_1$ calculations exclude values equal to zero (field data: 6 cases), $V_1$ calculations exclude values equal to zero (field data: 13 cases) since $V_1=0$ results in $y_s=0$ which is unreasonable; $\mu$=mean; $\sigma$=standard deviation; GEV= General Extreme Value Distribution; Gam = Gamma Distribution; LogNorm = Normal Logarithmic Distribution.

## 2. Methodology

This section explains the methodology used to assess the accuracy of the eight empirical equations analysed in this paper, which are listed in Table 1. The section also explains the methodology followed to explore the relative importance of the parameters of the scour equations via the two methods used in this work: OAT sensitivity analysis and a GSA.

2.1 Sources of data and data processing

Benedict & Caldwell (2014a, 2014b) assembled the scour database used in this study; this database is freely available online (Benedict & Caldwell 2014a, 2014b). Most (78%) of the data in Benedict & Caldwell (2014a, 2014b) were previously published in the National Bridge Scour Database of the U.S. Geological Survey (USGS), hence this paper refers to this dataset as the *USGS Database*. The database consists of 569 laboratory datapoints of local scour measurements around piers, and 936 local pier scour measurements from the field.

For the field data in the USGS Database includes a *MS Excel* spreadsheet of 28 abutment scour depth data, a *MS Excel* spreadsheet of 43 contraction scour depth data and a *MS Excel* spreadsheet of 507 local pier scour depth measurements. This research focuses on local pier scour; therefore, the abutment and contraction scour data were disregarded. The *USGS Database* also includes a *MS Access* spreadsheet with information about scour depth on piers at 93 sites. The *MS Access* spreadsheet includes information about the pier length and the measurement technique used for some on the piers which was is not included in the MS Excel spreadsheet. The pier scour depth data are complemented by reports that explain the data collection process.

For the laboratory data in the *USGS Database*, all the piers were considered circular, all the attack angles, $\theta$, were equal to zero and the pier length, $L$, was not provided by the *MS Excel* Spreadsheet. Moreover, the critical velocity, $V_c$ was provided for the 569 data in the database. The pier width, $B$, was available for all 569 data.

For the field data, the pier length, $L$, was unavailable for about 33% of the scour depth data (309/936). The unknown pier lengths, $L$, were estimated using a ratio based on the pier width, $B$: $L = 11.7B$, where 11.7 represents the average of the 627 known length-to-width ($L/B$) ratios in the database (Gavriel et al. 2023a). For all equations which use the pier length, $L$, to estimate scour depth, observations based on $L = 5.85B$ ($0.5 \times 11.7$) and $L = 23.4B$ ($2 \times 11.7$) were also carried out to examine how the scour depth estimation changes for each equation when the estimate of $L$ changes. The aforementioned assumptions were used to model unknown pier length data for use in the *CIRIA, HEC-18, Melville, Melville & Sutherland* and *Laursen* equations.

The *TAMU* equation by Briaud (2015a) requires the spacing of the piers, $S$, as an input for the estimation of scour depth. Information about the pier spacing, $S$, was not clearly detailed in the database reports, therefore the *TAMU* scour depth was calculated for multiple spacing values to examine how the spacing factor affects the scour estimation when it changes. It was observed that as spacing between the piers increased, scour depth stabilised (see also Gavriel et al. 2022). For the purposes of this analysis, *S* = 3m was used for estimates of spacing when using the *USGS Database* (field data) (for more details see Gavriel (2025)).

In this paper, results for calculated scour depth are subject to 10% error for both field and laboratory data. The 10% error is based on Figure 3.1 from Arneson et al. (2012) which shows that maximum scour is, as a rule of thumb, 10% larger than the maximum clear-water scour.

2.2 Equation Accuracy Assessment

Measured versus predicted scatter plots were utilized to provide a visual assessment of the accuracy of the eight scour equations in the field compared to the laboratory. Two boundary lines were added to the scatter plots: the ±50% boundary lines (Equation: $y = x \pm 0.5x$ ) and the factor of 1.5 boundary lines (Equation: $y = 1.5x;\ y = 1.5/x$). The percentage of datapoints that lie between the boundary lines was counted and used as a measure of the equation accuracy: the higher the percentage, the better the scour depth estimate of the empirical equation. The line of equality ($y = x$) was also drawn on each scatter plot to visually identify the number of points that were over- or underestimated by each equation.

2.3 Sensitivity Assessment: OAT Sensitivity Analysis

OAT sensitivity analysis was used to identify the most influential parameter in each equation. The analysis works by altering one parameter at the time while maintaining the other parameters constant to a baseline (e.g., Saltelli 2002; Ikonen 2016).

To apply the OAT sensitivity analysis, it was assumed that the parameters are normally distributed and the mean, *μ*, standard deviation, *σ*, *σ ± μ*, and *σ ± 0.5μ*, maximum and minimum are calculated for each parameter. The parameter ranges are shown in Table 3. Two base case scenarios were considered, one for the field data and one for the laboratory data. In each scenario, the baseline values of all parameters were taken as equal to their mean (Gavriel et al. 2023b).

To estimate the effect of each parameter on the scour depth estimation, Eq. 1 was used:

$$t_i = \frac{f(x_i)}{f(x_b)} \times 100 \qquad [1]$$

where, $t_s$ = spread, $f(x_i)$ calculated scour depth when the i-th parameter is set to a perturbed value (its minimum, or its maximum, or *μ - σ*, or *μ + σ*), $f(x_b)$ = calculated scour depth when all parameters are at baseline value. For each parameter, the difference between the maximum *$t_i$* value and the minimum *$t_i$* value across the four experiments was calculated using Eq. (2):

$$t_{i(\Delta)} = \max t_i - \min t_i \qquad [2]$$

For each equation the parameters are ranked based on the value of $t_{i(\Delta)}$. The higher the $t_{i(\Delta)}$, the higher the influence of the parameter uncertainty on the output.

## 2.4 Sensitivity assessment: GSA

As an alternative to OAT sensitivity analysis, this paper explores for the first time (to the authors' knowledge) in the context of scour equations the use of GSA (e.g. Saltelli et al 1999, Pianosi et al 2016). GSA essentially has four key steps. Step 1: The uncertainty in the model parameters is represented by probability distributions. Step 2: a set number of parameter combinations is generated by statistical sampling (e.g. Song & Kawai 2023) from the parameter distributions defined in Step 1. Step 3: the empirical equation is evaluated with each parameter combination, thus obtaining a synthetic data set of scour depth estimates. Step 4: the synthetic scour depth dataset is then analysed to derive a set of global sensitivity indices, one for each studied parameter, with each index measuring the relative contribution to the equation output uncertainty due to variations of that parameter. The definition of the sensitivity indices, and the calculation procedure to approximate their values from the output dataset, vary depending on the GSA method employed. In this work, the PAWN method by Pianosi & Wagener (2015) was used, which measures parameter sensitivity by looking at the difference between the unconditional output distribution when all parameters are varied simultaneously, and the conditional distribution when that parameter is fixed. Specifically, PAWN uses the Kolmogorov-Smirnov ($KS$) statistic to measure the maximum vertical distance between the unconditional Cumulative Distribution Functions (CDFs) of the output and several conditions CDFs at different conditioning values.

The parameter distributions for the GSA experiments were obtained by analysing the parameter values available in the *USGS database*. If only the minimum and maximum value of the parameter were known, its distribution was assumed as uniform over that min-max range. If instead the *USGS database* included a set of values for the parameter, a probability distribution was fitted to those values using the Akaike statistic (Akaike 1974) as best-fit measure. The resulting distributions fitted to the data for each parameter are detailed in Table 3.

Latin-Hypercube Sampling was then used to generate $N$ = 5000 parameter combinations for each empirical equation, and evaluated the equation against all the combinations, thus obtaining a sample of 5000 scour depth estimates.

The PAWN approximation strategy proposed by Pianosi & Wagener (2018) was used to calculate sensitivity indices. This requires first the calculation of the empirical distribution of all output (scour depth) samples – the unconditional CDF $F_y$. Then for each parameter its variability range was divided into $n$ = 10 intervals and the input-output dataset divided accordingly. Then the empirical distribution of the output was calculated within each of the 10 sub-samples – these are the 10 conditional CDFs $F_{y|xi}$. The PAWN sensitivity index was calculated as the mean of the Kolmogorov-Smirnov statistic (i.e. the maximum vertical difference) between each conditional CDF and the unconditional one (Eq. 3):

$$S_i = \operatorname*{mean}_{k=1,\ldots,n} KS(I_k) \quad \text{where} \quad KS(I_k) = \max |F_y(y) - F_{y|x_i}(y|x_i \in I_k)| \qquad [3]$$

The approach also allows for quantification of the uncertainty of the sensitivity indices due to using an input-output dataset of finite size. Specifically, lower and upper quantiles for each sensitivity index were calculated using bootstrapping, and the value of the sensitivity index corresponding to a "dummy" parameter was calculated via bootstrapping of the unconditional output distribution (Pianosi & Wagener 2018). By construction, this dummy sensitivity is purely the results of approximation errors and as such provides a reference value to assess the significance of all other indices: a sensitivity index value below the dummy sensitivity can be attributed to approximation errors rather than indicative of an actual contribution to the uncertainty of the output.

## 3. Results

### 3.1 Model accuracy

Figures 1 and 2 show the measured against the estimated scour depth of the *CIRIA* equation based on field data from the *USGS Database*. Figure 3 shows the same analysis from using the laboratory data. In all Figures (Figures S1 to S26), the more the points within the bounds, the more accurate the examined equation is. Similar plots for the other equations listed in Table 1 are shown in the *Supplementary Material* (Figures S1 to S26) and the accuracy results are summarized in Table 4.

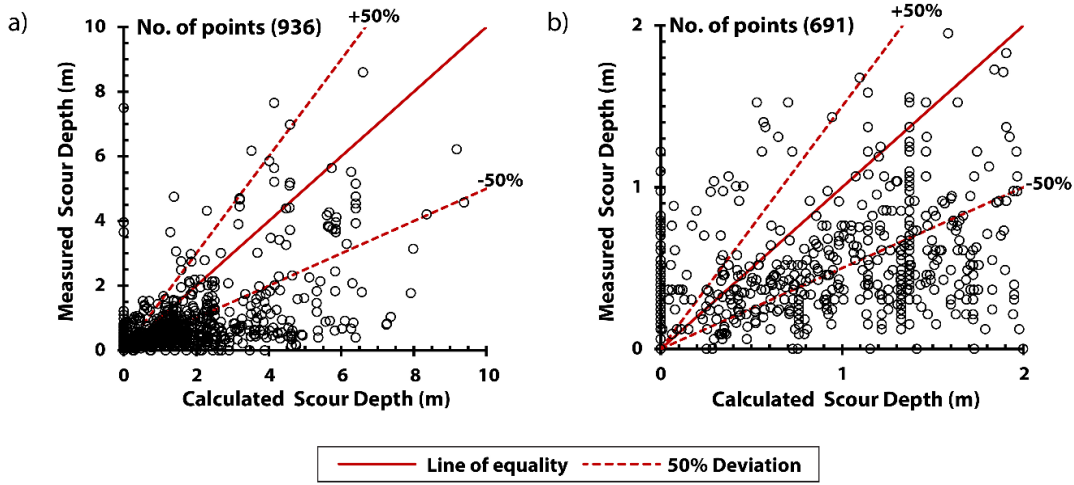

**Figure 1: (a) Measured scour depths from the *USGS Database* (field data) versus calculated values from the *CIRIA* equation, compared to the *y = x ± 0.5x* bounds; (b) same as (a) but focusing on measured scour depths ≤2m.**

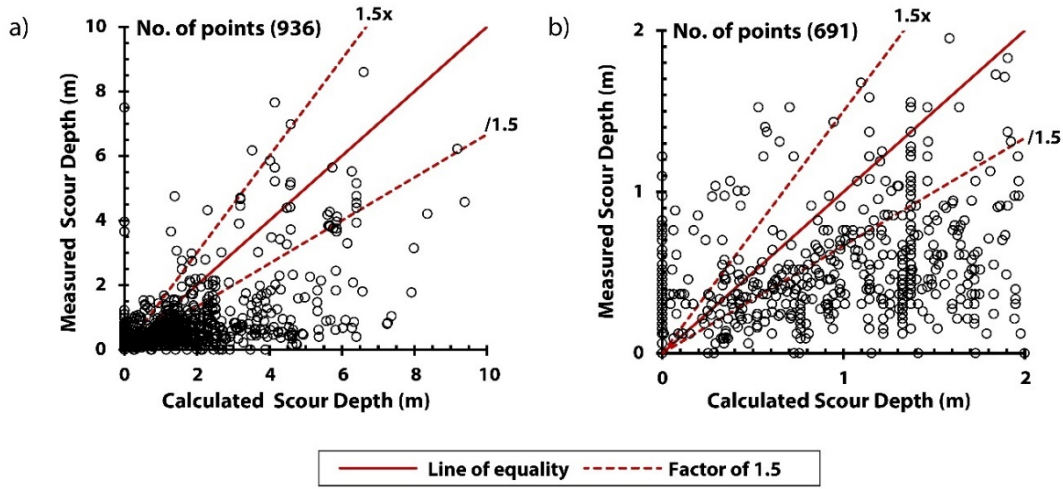

**Figure 2: (a)** Measured scour depths from the *USGS Database* (field data) versus calculated values from the *CIRIA* equation compared to *y* = 1.5*x* and *y* = *x* /1.5 bounds; **(b)** same as (a) but focusing on measured scour depths ≤ 2m

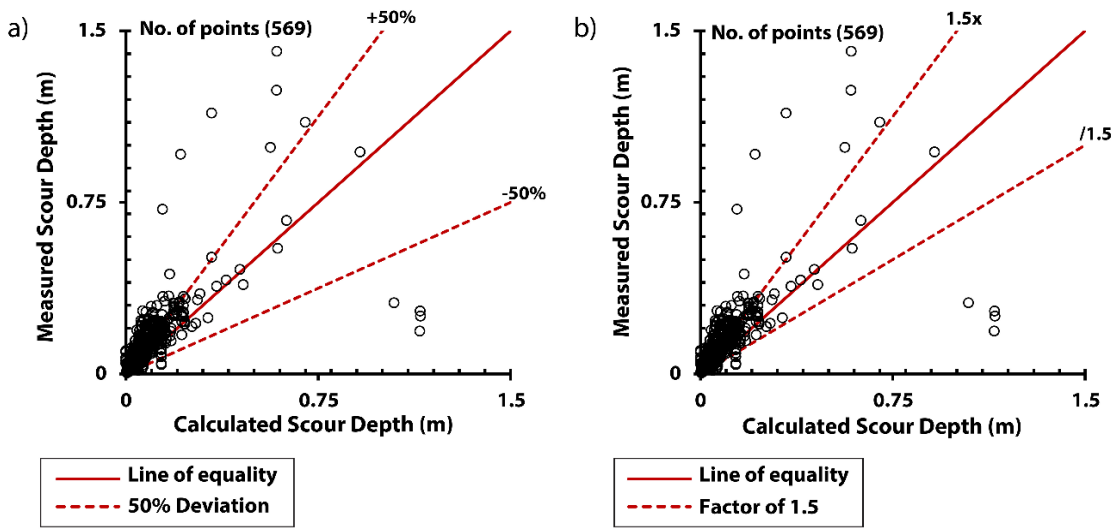

**Figure 3: (a)** Measured scour depths from the *USGS Database* (laboratory data) versus calculated values from the *CIRIA* equation, compared to *y* = *x* ± 0.5*x* bounds; **(b)** same as (a) but focusing on measured scour depths ≤ 2m

From Table 4, an overestimation trend is observed for most the studied equations. A higher concentration of data points within the bounds (i.e. higher accuracy) for results based on laboratory data than field data was identified, which is perhaps not surprising because laboratory data is collected in a more controlled environment. The *CIRIA* equation (which was partly derived from field data and party derived from laboratory data) was found to have the highest concentration of data within the ±50% and factor of 1.5 bounds, followed by the *Froehlich* equation.

The *HEC-18* equation has the highest concentration of data within the ±50% bounds for results based on laboratory data (83%), however, this level accuracy was not seen for the field data, for which only 22.1% of the data fall within the ±50% bounds. Similar reductions in accuracy between the analysis of the laboratory and the field data subsets were also observed for the *TAMU* and *Laursen* equations. The *Melville* and *Chitale* equations perform poorly for both the laboratory and field datasets with the percentage of points within the studied bounds lower than 10% in both conditions. For all the investigated equations, there is an increase in the percentage of data within the studied bounds when considering field data of with measured scour depths ≤ 2m.

3.2 OAT sensitivity analysis

Figure 4 shows the results from the OAT sensitivity analysis using the parameter ranges from the laboratory dataset for the eight empirical equations. Figure 5 shows the results using the parameter ranges from the field dataset. The computed sensitivity indices for all the studied equations are reported in the *Supplementary Material* (Tables S2 to S33).

Figure 4 (results based on the laboratory dataset) shows that the approach flow depth, $y_1$, was found to be highly influential for all eight equations. The approach flow velocity, $V_1$, was found to be highly influential for all the studied equations except for *Laursen*. The pier width, *B*, is highly influential for all the studied equations except for *Chitale*. For the *Froehlich*, *Melville & Sutherland* and *Melville* equations $D_{50}$ was found to be also influential. For all the equations studied there is at least one influential input related to the hydraulic characteristics, which is perhaps expected as the scour depth is the result of the action of flowing water (cf. Arneson et al. 2012).

Figure 5 (results based on field data) shows again that all equations are sensitive to at least one hydraulic parameter (i.e. $V_1$, $y_1$, $V_c$). Figure 5 also shows that all equations (except *Chitale*) are highly sensitive to the geometric parameters (i.e. *B*, *L* and *θ*). From Figures 4 and 5 the pier width, *B*, was found to be very important for the estimation of scour depth for all eight equations for both field and laboratory measurements. This is reasonable and consistent with sensitivity assessments from previous studies (e.g., Ekuje 2020; Shahriar et al. 2021a, 2021b and Kim et al. 2024). For all equations, the other parameters are also important and their (OAT) sensitivity is similar or only marginally smaller than that of the pier width *B*. In the next section, the results of a GSA used to investigate whether differences in the relative importance of the equation parameters can be detected are reported.

**Table 4:** Number (*n*) and percentage (%) of datapoints overestimated, underestimated, lying within bounds of ± 50% and within factor of 1.5 for the eight empirical equations studied, using laboratory and field data from the *USGS Database*. Highest *percentage* within bounds (i.e. highest accuracies) are highlighted in bold.

| Methodology | Underestimated | | Overestimated | | Within ±50% | | Within factor 1.5 | |
|---|---|---|---|---|---|---|---|---|
| | *n* | (%) | *n* | (%) | *n* | (%) | *n* | (%) |
| *Laboratory Data (n = 568)* | | | | | | | | |
| CIRIA | 498 | 87.5 | 71 | 12.5 | 277 | 48.7 | 267 | 46.9 |
| TAMU | 325 | 57.1 | 244 | 42.9 | 304 | 53.4 | 254 | 44.6 |
| HEC-18 | 138 | 24.3 | 431 | 75.7 | 472 | **83.0** | 402 | 70.7 |
| Melville | 541 | 95.1 | 28 | 4.90 | 51 | 9.00 | 42 | 7.4 |
| Froehlich | 52 | 9.1 | 517 | 90.9 | 157 | 27.6 | 83 | 14.6 |
| Melville & Sutherland | 306 | 53.8 | 263 | 46.2 | 119 | 20.9 | 64 | 11.2 |
| Chitale | 34 | 6.0 | 535 | 94.0 | 44 | 7.7 | 22 | 3.9 |
| Laursen | 105 | 18.5 | 464 | 81.5 | 464 | 81.5 | 416 | **73.1** |
| *Field Data (n = 936)* | | | | | | | | |
| CIRIA | 178 | 19.0 | 758 | 81.0 | 339 | **36.2** | 232 | **24.8** |
| TAMU | 113 | 12.1 | 732 | 78.2 | 142 | 15.2 | 87 | 9.3 |
| HEC-18 | 97 | 10.4 | 839 | 89.6 | 207 | 22.1 | 135 | 14.4 |
| Melville | 54 | 5.8 | 882 | 94.2 | 71 | 7.60 | 38 | 4.1 |
| Froehlich | 641 | 68.5 | 295 | 31.5 | 272 | 29.1 | 226 | 24.1 |
| Melville & Sutherland | 131 | 14.0 | 805 | 86.0 | 203 | 21.7 | 97 | 10.4 |
| Chitale | 17 | 1.8 | 919 | 98.2 | 44 | 4.7 | 22 | 2.4 |
| Laursen | 20 | 2.1 | 916 | 97.9 | 172 | 18.4 | 63 | 6.7 |
| *Field Data ≤2m* | | | | | | | | |
| CIRIA (*n = 619*) | 161 | 23.3 | 458 | 74.0 | 267 | **38.6** | 184 | **26.6** |
| TAMU (*n = 495*) | 204 | 41.2 | 291 | 58.8 | 119 | 24.0 | 75 | 15.2 |
| HEC-18 (*n = 548*) | 90 | 16.4 | 458 | 83.6 | 131 | 23.9 | 91 | 16.6 |
| Melville (*n = 512*) | 53 | 21.8 | 190 | 78.2 | 65 | 26.7 | 35 | 14.4 |
| Froehlich (*n = 916*) | 641 | 70 | 275 | 30.0 | 271 | 29.6 | 225 | 24.6 |
| Melville & Sutherland (*n = 512*) | 128 | 25 | 384 | 75.0 | 129 | 25.2 | 73 | 14.3 |
| Chitale (*n = 162*) | 17 | 10.6 | 144 | 89.4 | 37 | 23.0 | 21 | 13 |
| Laursen (*n = 446*) | 12 | 2.7 | 434 | 97.3 | 67 | 15.0 | 20 | 4.5 |

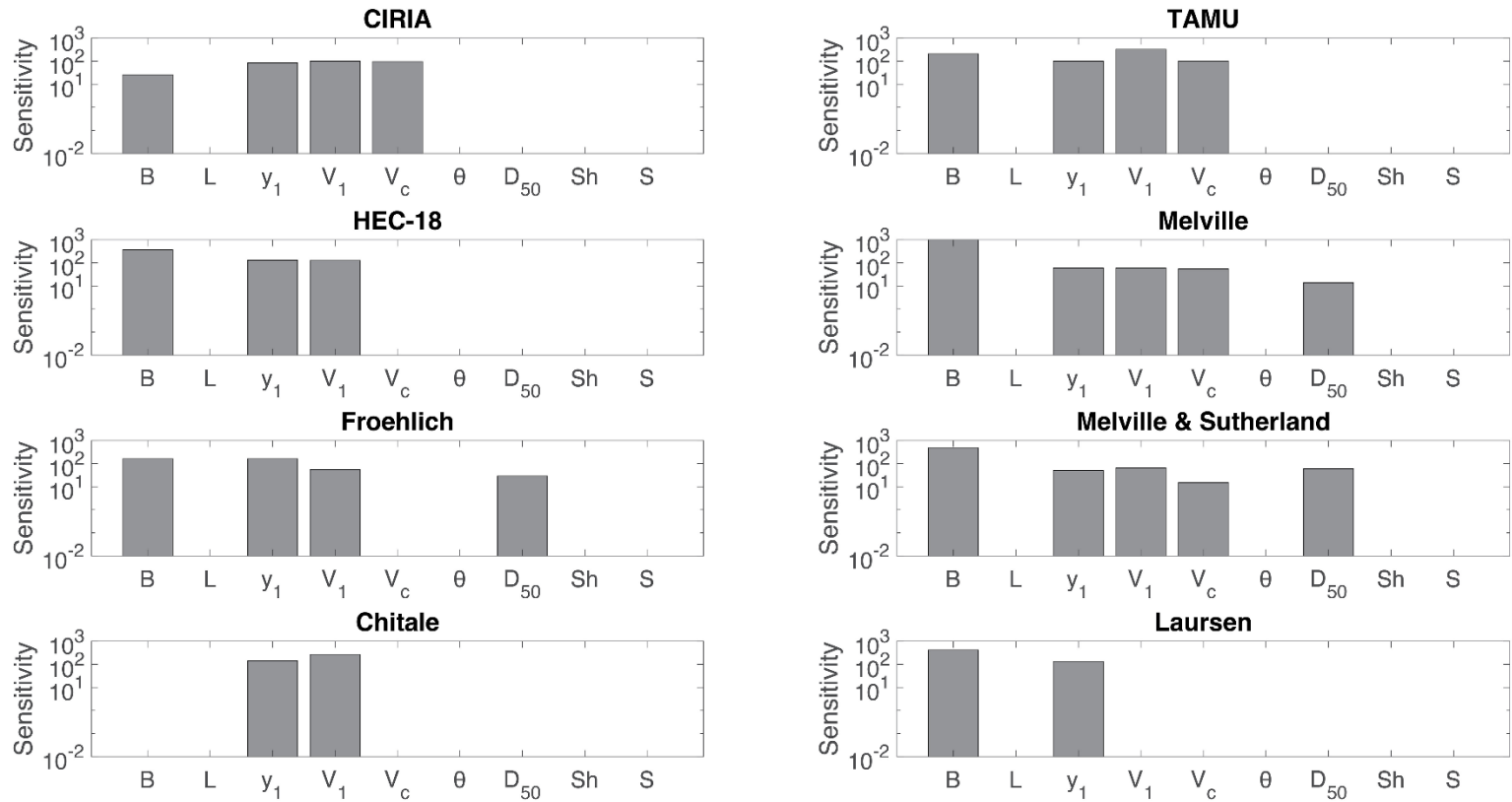

**Figure 4.** One-at-a-time sensitivity analysis results with parameter variations based on laboratory data. Each panel shows results for one of the eight empirical equations studied. The higher the value of the sensitivity index on the *y*-axis, the more influential the parameter on estimates of scour depth.

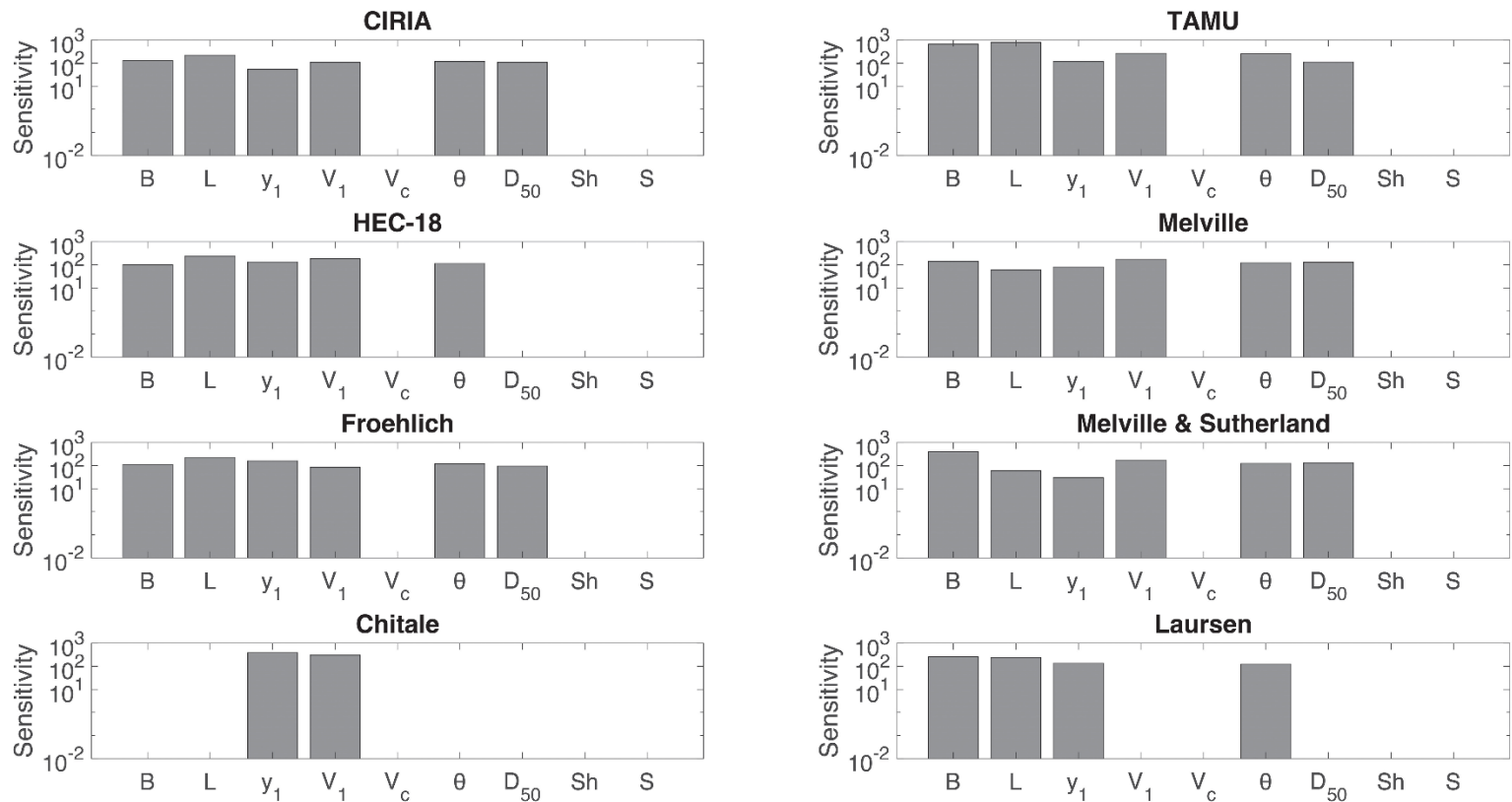

Figure 5. One-at-a-time sensitivity analysis results with parameter variations based on field data. Each panel shows results for one of the eight empirical equations studied. The higher the value of the sensitivity index on the *y*-axis, the more influential the parameter on estimates of scour depth.

### 3.3 Global Sensitivity Analysis

Figure 6 shows the PAWN sensitivity indices computed using Eq. (3) using the parameter distributions inferred from the laboratory dataset for the eight empirical equations listed in Table 1. Figure 7 shows the computed sensitivity indices when using the parameter distributions based on the field dataset. The dotted grey line on each of the plots in Figures 6 and 7 represents the PAWN sensitivity value for the dummy parameter.

Figure 6 (results based on laboratory data) shows that pier width, $B$, is the most influential parameter for *CIRIA*, *TAMU* and *Chitale*. The approach flow depth, $y_1$, appears to be the most influential parameter for the *HEC-18*, *Froehlich* and *Laursen* equations. The average riverbed grain size, $D_{50}$, and the approach velocity, $V_1$, are the most influential parameters for the *Melville* and *Melville & Sutherland* equations. In general, the parameter influences are much more varied than those from the OAT analysis (Figures 4), both across parameters for the same equation and across the set of studied equations.

Figure 7 (results based on field data) shows that the angle of attack, $\theta$, is the most influential parameter for the *TAMU*, *HEC-18*, *Melville & Sutherland* and *Chitale* equations. The pier shape, $Sh$, is highly influential for the *CIRIA*, *Melville*, *Froehlich*, *Melville & Sutherland* and *Chitale* equations. The approach flow depth, $y_1$, is either highly or moderately influential for all equations. The pier width, $B$, is highly influential only in the *Froehlich* equation. The approach velocity, $V_1$, is highly influential for the *Chitale* and *TAMU* equations.

Unlike the results from the OAT sensitivity analysis, results from the PAWN analysis show that some parameters have minor influence on the output. For example, pier width, $B$, and pier length, $L$, are found to have significant influence in the *HEC-18* equation based on the results of the OAT sensitivity analysis (Figure 5) but based on the results of the PAWN analysis (Figure 7) their influence on the output is negligible. Another example of disagreement between the two methodologies is the *Melville* equation, for which the parameters $B$, $V_1$ and $D_{50}$ are shown to have significant influence by the OAT analysis (Figure 5) and negligible influence by the PAWN analysis (Figure 7). Similar differences are found for the *CIRIA*, *TAMU*, *Froehlich* and *Melville & Sutherland* equations. The detailed comparison between OAT and PAWN sensitivity results is found in the *Supplementary Material* (Tables S34 and S35).

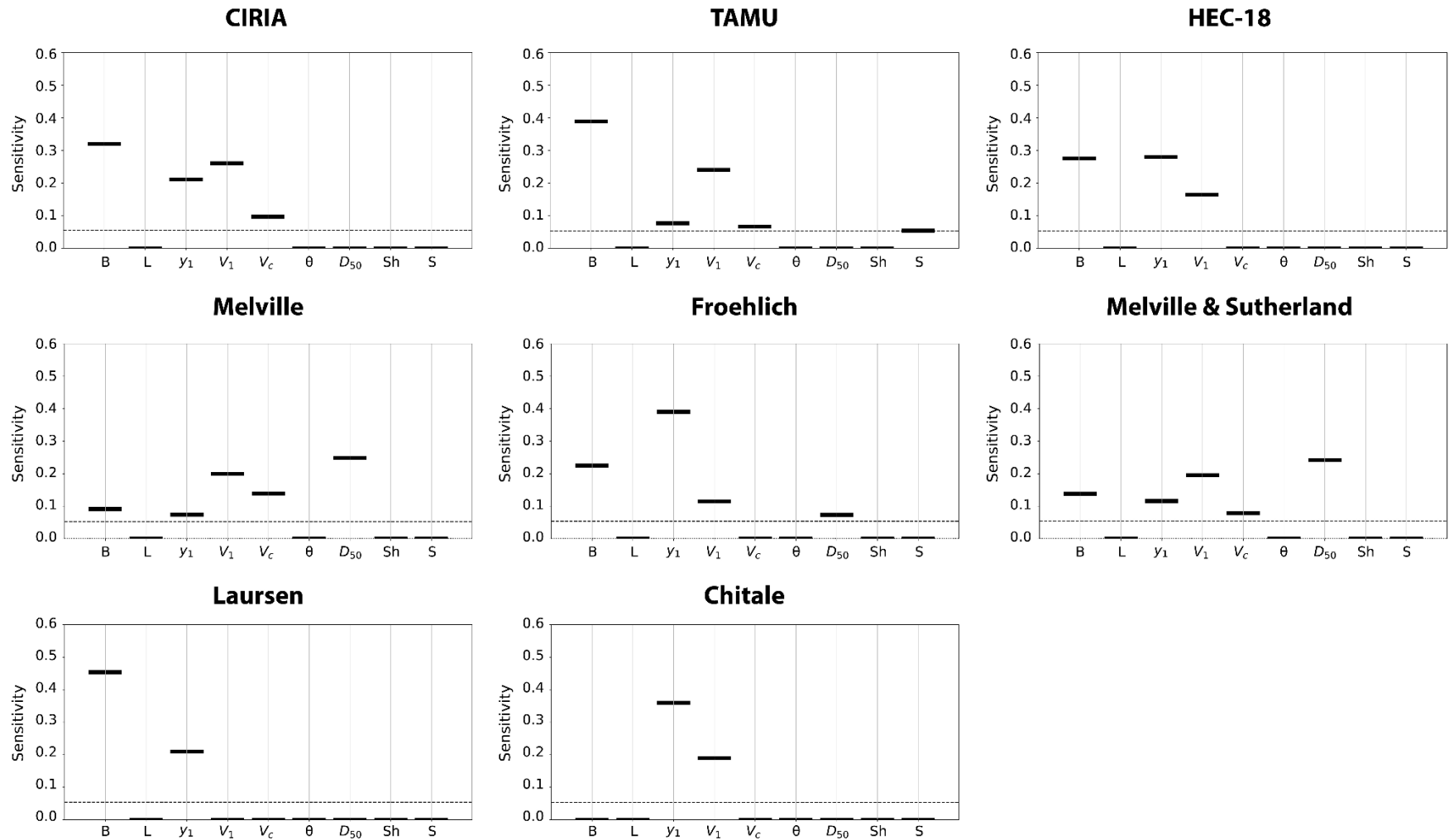

Figure 6. Global sensitivity analysis results using parameter distributions based on the laboratory dataset. Each panel shows results for one of the eight empirical equations. The higher the value of the PAWN sensitivity index (mean KS) on the $y$-axis, the more influential the parameter on estimates of scour depth.

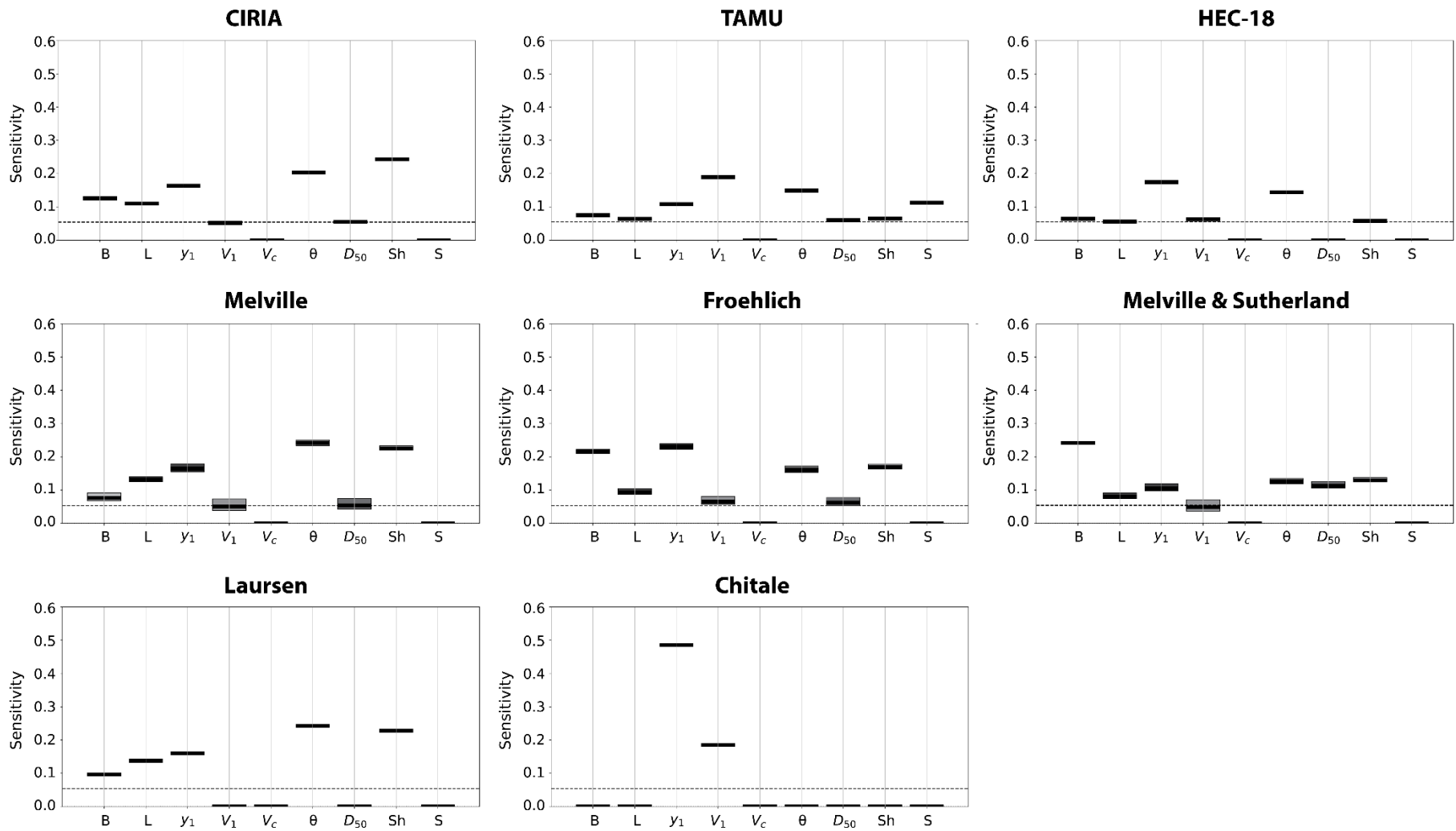

**Figure 7.** Global sensitivity analysis results using parameter distributions based on the field dataset. Each panel shows results for one of the eight empirical equations. The higher the value of the PAWN sensitivity index (mean KS) on the $y$-axis, the more influential the parameter on estimates of scour depth.

## 4. Summary and Conclusions

This study used the USGS field and laboratory database to evaluate the accuracy of eight empirical equations which can be used to predict pier scour depth. This study also applied two different types of sensitivity analysis (OAT and GSA) to the eight empirical equations to quantify the influence of the parameter uncertainty on scour predictions. This study showed that empirical equations perform very differently for laboratory and field data and highlighted the need for calibration of empirical equations with newly collected data from the field. This study also showed that using a GSA approach yields a better understanding of the influence of the parameters in the studied equations.

The application of model accuracy to the eight empirical equations showed that the good performance of some equations (i.e. *HEC-18*, *Laursen*) for predicting laboratory data of scour depth is not seen for prediction of scour depths in the field. This study argues that the equations calibrated using laboratory data do not capture the uncertainties from the field. Likewise, equations that were calibrated partly or entirely with field conditions, seem to perform better (but still with larger errors) for prediction of field scour depths. Specifically, the *CIRIA* equation and the *Froehlich* equation have the highest concentration of data within the ±50% and factor of 1.5 bounds (Table 4). Calibration of these equations larger field datasets may allow versions to be derived with improved accuracy.

This study has also shown that a GSA approach is easily applicable to bridge scour equations and provides a clearer picture of the relative importance of input parameters than a traditional approach such as OAT sensitivity analysis. Based on results from the GSA, the three most significant parameters for estimating scour in the field are angle of attack, pier shape and approach flow depth, which confirms the important role of an environmental factor (flow depth) on scour depths.

**CRediT authorship contribution statement**

Gianna Gavriel: *Conceptualisation, Methodology, Software, Formal Analysis, Data Curation, Investigation, Visualization, Writing – original draft, Writing – review & editing*. Maria Pregnolato: *Conceptualisation, Supervision*. Francesca Pianosi: *Conceptualisation, Methodology, Software, Writing – review & editing*. Theo Tryfonas: *Conceptualisation, Supervision*. Paul J. Vardanega: *Conceptualisation, Methodology, Writing – review & editing, Supervision, Project Administration*.

**Acknowledgements**

The first author thanks Dr F. Sarrazin, Dr V. Noacco and Mr T. Constantinides for their assistance with the implementation of the GSA code.

**References**

Akaike, H. (1974). A new look at the statistical model identification. *IEEE Transactions on Automatic Control*, **19(6):** 716-723. https://doi.org/10.1109/TAC.1974.1100705

Arneson, L.A., Zevenbergen, L.W., Lagasse, P.F. & Clopper, P.E. (2012). Evaluating scour at bridges (Fifth edition), *Hydraulic Engineering Circular No. 18. Federal Highway Administration, U.S. Department of Transportation*, Washington, DC, USA. Available from: < https://www.fhwa.dot.gov/engineering/hydraulics/pubs/hif12003.pdf > [08/01/2026].

Benedict, S.T. & Caldwell, A.W. (2014a). A pier-scour database—2,427 field and laboratory measurements of pier scour. *U.S. Geological Survey Data Series 845*. https://doi.org/10.3133/ds845

Benedict, S.T. & Caldwell, A.W. (2014b). A pier-scour database—2,427 field and laboratory measurements of pier scour: MS Excel database. *U.S. Geological Survey Data Series 845*. Available from: < https://pubs.usgs.gov/ds/0845/ > [08/01/2026].

Blench, T., Bradley, J.N., Joglekar, D.V., Bauer, W.J., Tison, L.J., Chitale, S.V., Thomas, A.R., Ahmad, M. & Romita, L.R. (1962). Discussion of "Scour at bridge crossings." *Transactions of the American Society of Civil Engineers*, **127(1):** 180-207. https://doi.org/10.1061/TACEAT.0008391

Brandimarte, L., Paron, P. & Di Baldassarre, G. (2012). Bridge pier scour: A review of processes, measurements, and estimates. *Environmental Engineering & Management Journal (EEMJ)*, **11(5):** 975–989.

Briaud, J-L. (2015a). Scour depth at bridges: Method including soil properties. I: Scour depth prediction. *Journal of Geotechnical and Geoenvironmental Engineering (ASCE)*, **141(2):** [04014104]. https://doi.org/10.1061/(ASCE)GT.1943-5606.0001222

Briaud, J-L. (2015b). Scour depth at bridges: Method including soil properties. II: Time rate of scour prediction. *Journal of Geotechnical and Geoenvironmental Engineering (ASCE)*, **141(2):** [04014105]. https://doi.org/10.1061/(ASCE)GT.1943-5606.0001223

Briaud, J-L., Gardoni, P. & Yao, C. (2014). Statistical, risk, and reliability analyses of bridge scour. *Journal of Geotechnical and Geoenvironmental Engineering (ASCE)*, **140(2):** [04013011]. https://doi.org/10.1061/(ASCE)GT.1943-5606.0000989

Choi, S-U. & Choi, B. (2016). Prediction of time-dependent local scour around bridge piers. *Water & Environment Journal*, **30(1–2):** 14–21. https://doi.org/10.1111/wej.12157

Dehghani, A.A., Esmaeili, T. & Chang, W-Y. & Dehghani, N. (2013). 3D numerical simulation of local scouring under hydrographs. *Water Management*, **166(3):** 120–131. https://doi.org/10.1680/wama.11.00043

Deng, L. & Cai, C.S. (2010). Bridge Scour: Prediction, Modelling, Monitoring, and Countermeasures—Review. *Practice Periodical on Structural Design and Construction (ASCE)*, **15(2):** 125–134. https://doi.org/10.1061/(ASCE)SC.1943-5576.0000041

Dikanski, H., Imam, B. & Hagen-Zanker, A. (2018). Effects of uncertain asset stock data on the assessment of climate change risks: a case study of bridge scour in the UK. *Structural Safety*, **71:** 1–12. https://doi.org/10.1016/j.strusafe.2017.10.008

Ekuje, F.T. (2020). Bridge scour: climate change effects and modelling uncertainties. *Ph.D. thesis,* University of Surrey, Surrey, UK. https://doi.org/10.15126/thesis.00849796

Ettema, R., Constantinescu, G. & Melville, B. (2017). Flow-Field Complexity and Design Estimation of Pier-Scour Depth: Sixty Years since Laursen and Toch. *Journal of Hydraulic Engineering (ASCE)*, **143(9):** [03117006]. https://doi.org/10.1061/(ASCE)HY.1943-7900.0001330

Froehlich, D.C. (1988). Analysis of On-Site Measurements of Scour at Piers. In: *Hydraulic Engineering* (Abt, S.R. & Gessler, J. (eds.)), New York: American Society of Civil Engineers, New York, pp. 534–539.

Gaudio, R., Grimaldi, C., Tafarojnoruz, A. & Calomino, F. (2010). Comparison of Formulae for the Prediction of Scour Depth at Piers. In: *Proceedings of the First European IAHR Congress (Edinburgh, 2010).* Available from:
< https://www.iahr.org/library/infor?pid=10899 > [08/01/2026].

Gavriel, G. (2025). Evaluating Parameter Uncertainty of Empirical Equations for Assessing Local Scour around Bridge Piers. *Ph.D. thesis*, University of Bristol, Bristol, UK.

Gavriel, G., Yaqoobi, W., Pregnolato, M., Tryfonas, T. & Vardanega, P.J. (2025). A study on empirical prediction and monitoring of bridge pier scour depth: implications for asset owners and operators. In: *Proceedings of the 5th International Conference on Evolving Cities (2025)* (Bahaj, A.S. (ed.)), Evolving Cities Publications, Southampton, UK, article number: 1151, 8pp. https://doi.org/10.55066/proc-icec.2025.1151

Gavriel, G., Vardanega, P.J. & Pregnolato, M. (2023a). Comparison of maximum scour estimations using the HEC-18 method with field data from the USGS database. In: *European Geotechnical Engineering - Unity and Diversity: Proceedings of the 17th Danube–European Conference on Geotechnical Engineering, 7-9 June 2023, Bucharest, Romania* (Lungu, I., Teodoru, I-B. & Batali, L. (eds.)). Politehnium Publishing House, Işai, România, vol. 2, pp. 631-638.

Gavriel, G., Pregnolato, M. & Vardanega, P.J. (2023b). Using the USGS database to study parameter uncertainty when assessing pier scour using the HEC-18 framework. In: *Life-Cycle of Structures and Infrastructure Systems: Proceedings of the Eighth International Symposium on Life-Cycle Civil Engineering (IALCCE 2023), 2-6 July 2023, Politecnico Di Milano, Milan, Italy* (Biondini, F. & Frangopol, D.M. (eds.)), CRC Press/Balkema (Taylor & Francis Group), Abingdon, Oxon, UK, pp. 3230-3237. https://doi.org/10.1201/9781003323020-394

Gavriel, G., Vardanega, P.J. & Pregnolato, M. (2022). Preliminary comparison of scour depth estimation methods. In: *Bridge Safety, Maintenance, Management, Life-Cycle, Resilience and Sustainability: Proceedings of the Eleventh International Conference on Bridge Maintenance, Safety and Management (IABMAS 2022), Barcelona, Spain, July 11-15, 2022* (Casas, J.R., Frangopol, D.M. & Turmo, J. (eds.)), CRC Press/Balkema (Taylor & Francis Group), London, UK, pp. 2107-2113. https://doi.org/10.1201/9781003322641-261

Graf, W.H. & Istiarto, I. (2002). Flow pattern in the scour hole around a cylinder. *Journal of Hydraulic Research*, **40(1):** 13–20. https://doi.org/10.1080/00221680209499869

Hong, J-H., Goyal, M. K., Chiew, Y-M. & Chua, L.H.C. (2012). Predicting time-dependent pier scour depth with support vector regression. *Journal of Hydrology*, **468–469:** 241–248. https://doi.org/10.1016/j.jhydrol.2012.08.038

Ikonen, T. (2016). Comparison of global sensitivity analysis methods – Application to fuel behaviour modelling. *Nuclear Engineering & Design*, **297:** 72-80. https://doi.org/10.1016/j.nucengdes.2015.11.025

Kazemian, A., Yee, T., Oguzmert, M., Amirgholy, M., Yang, J. & Goff, D. (2023). A review of bridge scour monitoring techniques and developments in vibration-based scour monitoring for bridge foundations. *Advances in Bridge Engineering*, **4:** [2]. https://doi.org/10.1186/s43251-023-00081-6

Kim, T., Shahriar, A.R., Lee, W-D. & Gabr, M.A. (2024). Interpretable machine learning scheme for predicting bridge pier scour depth. *Computers and Geotechnics*, **170:** [106302]. https://doi.org/10.1016/j.compgeo.2024.106302

Kirby, A.M., Roca, M., Kitchen, A., Escarameia, M. & Chesterton, O.J. (2015). Manual on scour at bridges and other hydraulic structures, second edition. *Report no. CIRIA C742*, Department of Transport, London, UK.

Kumar, V., Baranwal, A. & Das, B.S. (2024). Prediction of local scour depth around bridge piers: modelling based on machine learning approaches. *Engineering Research Express*, **6(1):** [015009]. https://doi.org/10.1088/2631-8695/ad08ff

Lauchlan, C. & May, R. (2002). Comparison of General Scour Prediction Equations for River Crossings. In: *Proceedings of ICSF-1 First International Conference on Scour of Foundations* (Chen, H-C. & Briaud, J-L. (eds.)), Texas Transportation Institute, Publications Department, pp. 184–197. Available from: < https://hdl.handle.net/20.500.11970/100333 > [08/01/2026].

Laursen, E.M. & Toch, A. (1956). Scour Around Bridge Piers And Abutments. Iowa Institute of Hydraulic Research, State University of Iowa, Iowa, USA. Available from: < https://publications.iowa.gov/20237/1/IADOT_hr_30_bulletin_4_Scour_Bridge_Piers_Abutments_1956.pdf > [08/01/2026].

Laursen, E.M. (1962). Scour at Bridge Crossings. *Transactions of the American Society of Civil Engineers (ASCE)*, **127(1):** 166-180. https://doi.org/10.1061/TACEAT.0008432

Liang, F., Zhang, H. & Huang, M. (2015). Extreme scour effects on the buckling of bridge piles considering the stress history of soft clay. *Natural Hazards*, **77(2):** 1143–1159. https://doi.org/10.1007/s11069-015-1647-4

Maddison, B. (2012). Scour failure of bridges. *Proceedings of the Institution of Civil Engineers – Forensic Engineering*, **165(1):** 39–52. https://doi.org/10.1680/feng.2012.165.1.39

Melville, B.W. & Sutherland, A.J. (1988). Design Method for Local Scour at Bridge Piers. *Journal of Hydraulic Engineering (ASCE)*, **114(10):** 1210-1226. https://doi.org/10.1061/(ASCE)0733-9429(1988)114:10(1210)

Melville, B.W. (1997). Pier and Abutment Scour: Integrated Approach. *Journal of Hydraulic Engineering (ASCE)*, **123(2):** 125–136. https://doi.org/10.1061/(ASCE)0733-9429(1997)123:2(125)

Melville, B.W. & Chiew, Y-M. (1999). Time scale for local scour at bridge piers. *Journal of Hydraulic Engineering (ASCE)*, **125(1):** 59–65. https://doi.org/10.1061/(ASCE)0733-9429(1999)125:1(59)

Melville, B.W. & Coleman, S.E. (2000). *Bridge Scour.* Water Resources Publications LLC, Highlands Ranch, Colorado, USA.

Nandi, B. & Das, S. (2023). Identify most promising temporal scour depth formula for circular piers proposed over last six decades. *Ocean Engineering*, **286(Part2):** [115639]. https://doi.org/10.1016/j.oceaneng.2023.115639

National Academies of Sciences, Engineering, and Medicine. (2011a). *Evaluation of Bridge Scour Research: Pier Scour Processes and Predictions*. The National Academies Press, Washington, DC. https://doi.org/10.17226/22886

National Academies of Sciences, Engineering, and Medicine. (2011b). *Scour at Wide Piers and Long Skewed Piers*. The National Academies Press, Washington, DC. https://doi.org/10.17226/14426

Pianosi, F. & Wagener, T. (2015). A simple and efficient method for global sensitivity analysis based on cumulative distribution functions. *Environmental Modelling & Software*, **67:** 1–11. https://doi.org/10.1016/j.envsoft.2015.01.004

Pianosi, F. & Wagener, T. (2018). Distribution-based sensitivity analysis from a generic input-output sample. *Environmental Modelling & Software*, **108:** 197-207. https://doi.org/10.1016/j.envsoft.2018.07.019

Pianosi, F., Beven, K., Freer, J., Hall, J.W., Rougier, J., Stephenson, D.B. & Wagener, T. (2016). Sensitivity analysis of environmental models: A systematic review with practical workflow. *Environmental Modelling & Software,* **79:** 214-232. https://doi.org/10.1016/j.envsoft.2016.02.008

Qi, M., Li, J. & Chen, Q. (2016). Comparison of existing equations for local scour at bridge piers: parameter influence and validation. *Natural Hazards*, **82(3):** 2089–2105. https://doi.org/10.1007/s11069-016-2287-z

Richardson, E.V. & Davis, S.R. (2001). Evaluating Scour at Bridges: Fourth Edition. *Report No. FHWA NHI 01-001 HEC-18*, Federal Highway Administration, U.S. Department of Transportation, Washington, DC, USA. Available from: < https://www.engr.colostate.edu/CIVE510/Manuals/HEC-18%204th%20Ed.(2001)%20-%20Evaluating%20Scour%20at%20Bridges.pdf > [08/01/2026].

Saltelli, A., Aleksankina, K., Becker, W., Fennell, P., Ferretti, F., Holst, N., Li, S. & Wu, Q. (2019). Why so many published sensitivity analyses are false: A systematic review of sensitivity analysis practices. *Environmental Modelling & Software*, **114:** 29-39. https://doi.org/10.1016/j.envsoft.2019.01.012

Saltelli, A. (2002). Making best use of model evaluations to compute sensitivity indices. *Computer Physics Communications*, **145(2):** 280–297. https://doi.org/10.1016/S0010-4655(02)00280-1

Saltelli, A., Tarantola, S. & Chan, K.P.-S. (1999). A Quantitative Model-Independent Method for Global Sensitivity Analysis of Model Output. *Technometrics*, **41(1):** 39-56. https://doi.org/10.1080/00401706.1999.10485594

Shahriar, A.R., Ortiz, A.C., Montoya, B.M. & Gabr, M.A. (2021a). Bridge Pier Scour: An overview of factors affecting the phenomenon and comparative evaluation of selected models. *Transportation Geotechnics*, **28:** [100549]. https://doi.org/10.1016/j.trgeo.2021.100549

Shahriar, A.R., Montoya, B.M., Ortiz, A.C. & Gabr, M.A. (2021b). Quantifying probability of descendance estimates of clear water local scour around bridge piers. *Journal of Hydrology*, **597:** [126177]. https://doi.org/10.1016/j.jhydrol.2021.126177

Song, C. & Kawai, R. (2023). Monte Carlo and variance reduction methods for structural reliability analysis: A comprehensive review. *Probabilistic Engineering Mechanics*, **73:** [103479]. https://doi.org/10.1016/j.probengmech.2023.103479

Sun, Z., Dong, H., Sun, Y. & Li, Z. (2023). New formula for scour depth at piles based on energy equilibrium. *Ocean Engineering*, **287:** [115725]. https://doi.org/10.1016/j.oceaneng.2023.115725

Vardanega, P.J., Gavriel, G. & Pregnolato, M. (2021). Assessing the suitability of bridge-scour-monitoring devices. *Proceedings of the Institution of Civil Engineers – Forensic Engineering*, **174(4):** 105–117. https://doi.org/10.1680/jfoen.20.00022

Yazdanfar, Z., Lester, D., Robert, D. & Setunge, S. (2021). A novel CFD-DEM upscaling method for prediction of scour under live-bed conditions. *Ocean Engineering*, **220:** [108442]. https://doi.org/10.1016/j.oceaneng.2020.108442